\documentclass[aps,twocolumn]{revtex4}   %%preprint
\newcommand{\dzz}{\ensuremath{d_{z^2}}}
\newcommand{\dxxyy}{\ensuremath{d_{x^2-y^2}}}

\newcommand{\dxy}{\ensuremath{d_{xy}}}

\newcommand{\dxz}{\ensuremath{d_{zx}}}
\newcommand{\dyz}{\ensuremath{d_{yz}}}

\usepackage{amsmath}
\usepackage{color}   %JJ
\usepackage{bm}
\usepackage{url,amssymb,amsmath,graphics,graphicx,epsfig}
\usepackage{enumerate}
\usepackage{hyperref}
\usepackage{fancyvrb}
\begin{document}

\title{RIXS and excited states calculation with the Xclaim code}
%\author{Javier Fern\'{a}ndez-Rodr\'{\i}guez$^{1,2}$, Michel van Veenendaal$^{1,2}$ and Brian Toby$^{2}$\\ 
%$^1$  Department of Physics, Northern Illinois University, De Kalb, Illinois 60115, USA \\
%$^2$  Advanced Photon Source, Argonne National Laboratory, 9700 South Cass Avenue, Argonne, Illinois 60439, USA}

\author{Javier Fern\'{a}ndez-Rodr\'{\i}guez}
\author{Brian Toby}
\author{Michel van Veenendaal}

%%%%%%%%%%%%%%%%%%  NO HAY ADDRESS EN LA STANDARD ARTICLE DOCUMENT CLASS
\affiliation{Department of Physics, Northern Illinois University, DeKalb, Illinois 60115, USA}
\affiliation{Advanced Photon Source, Argonne National Laboratory, 9700 South Cass Avenue, Argonne, Illinois 60439, USA}

\date{\today}

\begin{abstract}
We discuss the calculation of resonant inelastic x-ray scattering
(RIXS) with Xclaim, a multi-plaform code for the calculation of
core-hole spectroscopy based on a ligand-field multiplet model.  RIXS
can be calcuated with a ligand-field model with
ligands for the L$_{2,3}$, M$_{2,3}$, $K_{\alpha}$ and $K_{\beta}$
edges for $d$-valence metals.  In addition, the program brings the possibility of fully
diagonalizing the hamiltonian to classify the excited states.
The command line use of xclaim is described for scripting purposes to
help in automatizing the results that can be obtained
with the graphical interface.
\end{abstract}

\maketitle

%\section{Introduction}

Resonant inelastic x-ray scattering (RIXS) is an experimental technique
for observing different types of excitations (dd, charge transfer,
magnons, etc. ) that has attracted a lot of
interest.~\cite{Kotani2001RMP,vanVeenendaalPRL2006}
Amongst its advantages can be cited chemical and bulk selectivity, and the possibility to be used at high pressure.
However, multiplet effects
can make its interpretation not straightforward.  A many-body model is necessary many times to 
take into account the character of the electronic structure of strongly correlated materials,
and the interaction with the core hole in the final states.

%Polarization dependence of RIXS
%% CITAS 1,2,3,review kotani DE RIXS de MvV PRL 2006
 %\cite{Wray2012}

%\section{The xclaim code}

\begin{figure}
\begin{center}
\includegraphics[width=0.95\columnwidth]{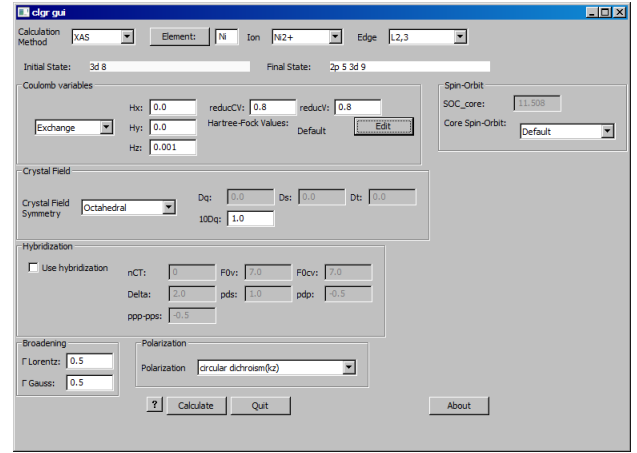}
\end{center}
\caption{\label{FigMainWindow} Main input window for xclaim.  Atomic, crystal field, and charge transfer parameters for the
calculation can be set in the input boxes.  Pop-up windows can be used for the detailed modification of Coulomb and spin-orbit parameters,
and the use of a general crystal-field in terms of Wybourne parameters.}  
\end{figure}

Xclaim~\cite{xclaimURL,xclaimPAPER} is a code for the calculation of core-hole spectroscopy using a
ligand-field multiplet model~\cite{vanVeenendaal2015book,deGrootKotani}. The
hamiltonian matrices and spectroscopy is calculated by a compiled
fortran code with the input parameters being set in a graphical interface in
python (Fig.~\ref{FigMainWindow}).  In this paper we
descibe the extension of the code for calculating RIXS spectra and classifying
the excited states.  We also discuss the use of xclaim as a command line tool,
that allows to reproduce previous calculations and automatize
the process of calculating spectra.

%%%%\begin{figure*}
%%%%\end{figure*}

\section{Calculation of RIXS spectra}

In the RIXS technique incoming x-ray photons ($\hbar\omega$) are tuned in the vicinity of an absorption edge of the resonant ion
and one measures the scattered intensity at for outgoing photons of energy  $\hbar\omega'$.
The RIXS cross-section~\cite{Ament2011RMP} can be calculated through the Kramers-Heisenberg formula
\begin{eqnarray}
\label{KramersHeisenberg}
I(\omega)&=&  \sum_{|f\rangle}   \langle f |T'^{\dagger}
\frac{1}{E_g + \hbar\omega   - H_i +i\Gamma}  T|g \rangle   \nonumber \\
&& \delta (   E_f  +\hbar\omega' - E_g - \hbar\omega ) .
\end{eqnarray}
% | i \rangle \langle i |
%
$|g\rangle$ corresponds to the ground state of the initial state
hamiltonian (without core-hole) and $|f\rangle$ labels the final states.
The intermediate denominator corresponds to the Green's function of the intermediate state hamiltonian $H_i$
with $\Gamma$ is the intermediate core-hole lifetime broadening
and the final $\delta$ function imposes energy conservation.
$T$ and $T'$ are the transition operators for the incoming and outgoing light,
$T=\bm{\epsilon}\mathbf{r}$ and $T'=\bm{\epsilon'}\mathbf{r}$ in the dipolar case for
incoming and outgoing beams with polarization vectors $\mathbf{\epsilon}$ and $\mathbf{\epsilon'}$

A common scattering geometry is to use 90$^o$ between incident and
scatterd beam to minimize the elastic scattering, and to use $\sigma$
and $\pi$ polarization for the incoming light.  No polarization
analysis is normally used for the outgoing beam, since it would
atenuate too much the beam.

%ARPACK

\begin{figure}
\begin{center}
\includegraphics[width=0.95\columnwidth]{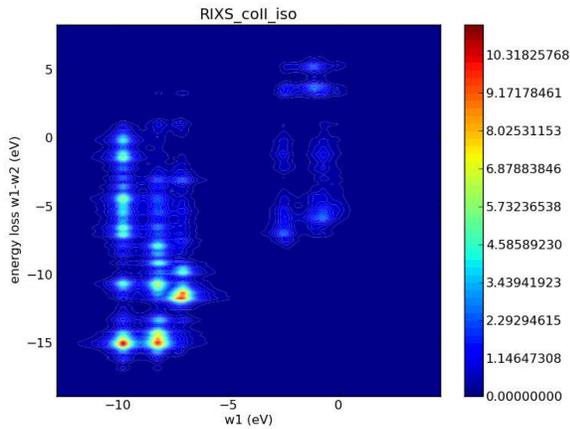}
\end{center}
\caption{\label{FigRIXS} Output contour plot for a RIXS calculation for Co$^{2+}$ at the K$_{\alpha}$ edge
based on a crystal field model~\cite{Ding2012PRB}.}
\end{figure}

%\subsection{Parallelization}

The RIXS calculation done by full diagonalizations of the hamiltonian matrices
can be subdivided into several independent processes.  
Each diagonalization can be separated into two independent processes for the upper and lower eigenvalues
(in the case of $K_\alpha$ and $K_\beta$ RIXS three hamiltonian matrices need to be diagonalized).
The lorentzian and gaussian convolutions of the RIXS map are also paralellized.
The usage of the multiprocessing python library allows to launch several independent python processes
from the python interpreter and doing the calculation in a computer
with several processor cores will notably speed up the RIXS calculation.

\section{Excited states}

\begin{figure}
\begin{center}
\includegraphics[width=0.99\columnwidth]{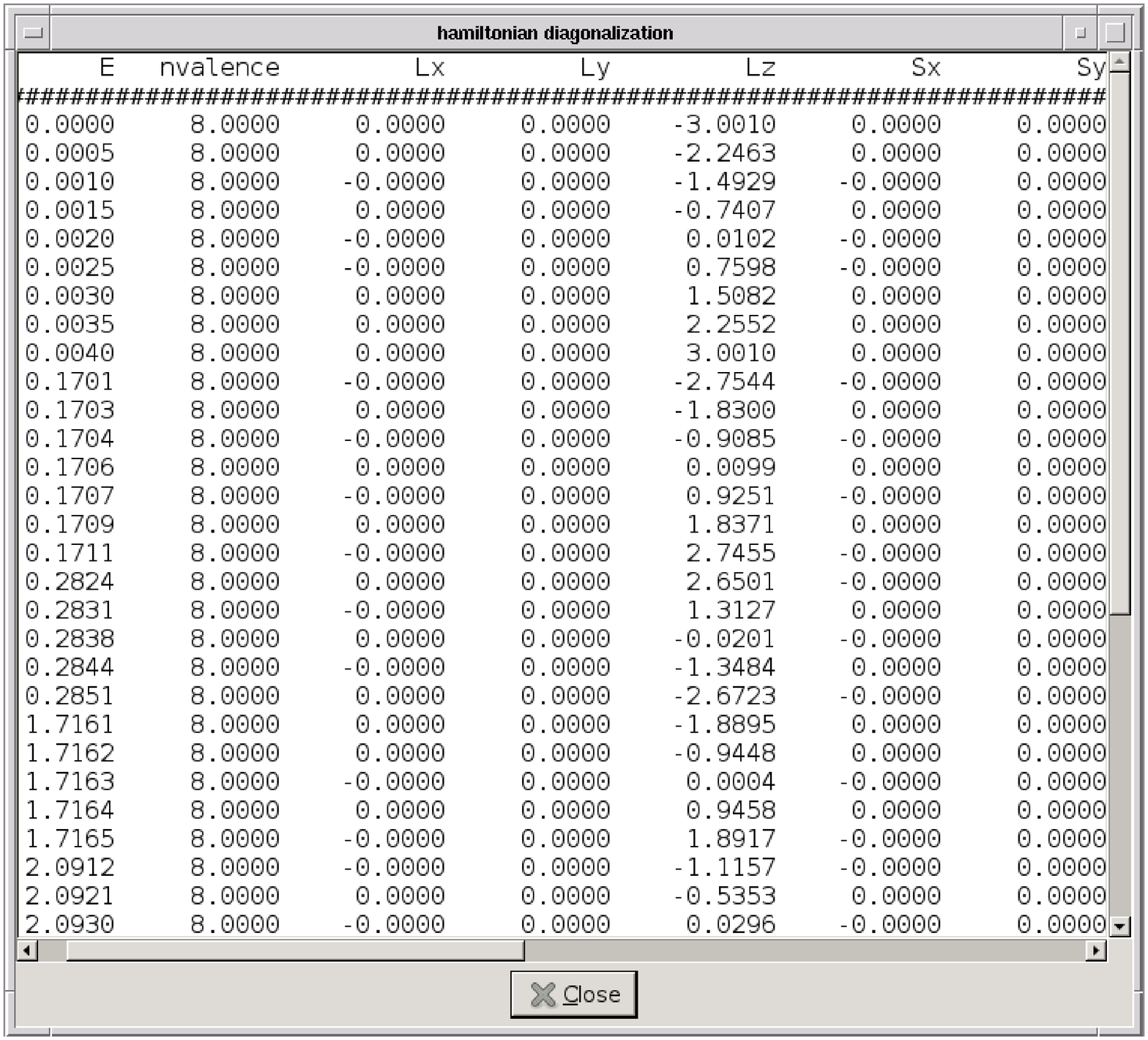}
\end{center}
\caption{\label{FigDiagonalize} Window showing the results of diagonalizing the hamiltonian.
}
\end{figure}

In addition to getting information from the ground state it is possible to also
characterize the excited states by doing an exact diagonalization of the
hamiltonian.
By selecting {\it diagonalize} in the spectroscopy tab the program
will fully diagonalize the hamiltonian and show a text window
with the energy positioning and information for the
excited states (Fig.~\ref{FigDiagonalize}).  The text window shows for each quantum state
the energy positioning $E$, the number of valence electrons, $x$, $y$ and $z$ components
of the orbital angular momentum $\langle\hat{\mathbf{L}}\rangle$ and spin $\langle\hat{\mathbf{S}}\rangle$,
the expectation value of the spin orbit operator for each state (labeled by $L.S$), the total spin
$\langle\hat{\mathbf{S}}^2\rangle=S(S+1)$ (this is related to the spin multiplicity, i.e. singlet, doublet$\ldots$), and the 
occupation of the different d orbitals: $\dzz$, $\dxxyy$, $\dxy$, $\dyz$, and $\dxz$.
In the case of a calculation with ligands it will output the occupation of the valence and ligand shells.

%%%jfrNiTPP,

Information on excitations can be used in classifying the features of RIXS spectra as different kind of $dd$~excitations
or in the interpretation of pump-probe experiments,
where the system undergoes a cascading decay through excited quantum states.~\cite{Gawelda2007,JunChang2010}.
Another interest for studying excitations
comes from finding systems where crystal-field exctiations very close
to the ground state allowing an easy switch the ground state by small changes in the ligand environment.~\cite{PRBFePcJFR}

%\begin{itemize}
%\item The information shown on Crystal-field (dd) and charge-transfer excitations
%\item transition-metal complex
%\end{itemize}

%\begin{figure*}
%\end{figure*}

\section{Command Line interface}

In addition to using the graphical interface, 
it is possible to run the program from the command-line and to repeat the calculation with the same parameters
as set previously on the graphical interface, or modifying them.
The file {\it out/input.txt} contains the parameters saved from the last execution of the
program.  It is worth noting that {\it out/input.txt} is rewritten each time the
{\it Calculate} button is pressed or when the command line interface
is run.
The input file is saved as a python dictionary, between braces and different lines formated as
{\it "label": value,} pairs.  Values can be numerical or a string given between quotes.
The {\it out} directory also contains other files depending on the type of calculation that is run:
{\it out/GS\_info.txt}  (calculated expectation values for the ground state)
and the files {\it out/poles\_*.txt}  (calulated spectra for different polarizations).

To use the comand line interface just type {\it cp out/input.txt\: input1.txt\: ; \: ./xclaim\_gui.py\: -i input1.txt}
The resulting spectra and ground state expectation values will be in
the directory {\it out} (its contents will be overwritten).
A complete list of options can be seen by typing {\it ./xclaim\_gui.py\: - -help}
The xclaim webpage~\cite{xclaimURL} contains more detailed information about the command line and rebroadening spectra.

%\section{Conclusions}

\section{Acknowledgements}

This work was supported by the U. S. Department
of Energy (DOE), Office of Basic Energy Sciences, Division of Materials
Sciences and Engineering under Award No. DE-FG02-03ER46097, the
time-dependent x-ray spectroscopy collaboration as part of the
Computational Materials Science and Chemistry Network
(CMSCN) under grant DE-FG02-08ER46540, and NIU Institute for
Nanoscience, Engineering, and Technology. Work at Argonne National
Laboratory was supported by the U. S. DOE, Office of Science, Office of
Basic Energy Sciences, under contract No. DE-AC02-06CH11357.

%\appendix
%\section{}
%%%%%%%%%%%%%%%%%%%%%%%%%%%%%%%%%%%%%%%%%%%%%%%%%%%%%%%%%%%%%%%%%%%%
%%%%%%%%%%%%%%%%%%%%%%%%%%%%%%%%%%%%%%%%%%%%%%%%%%%%%%%%%%%%%%%%%%%%
\bibliography{ManuscriptRIXS}
\end{document}